# Thermal conduction across a boron nitride and silicon oxide interface

Xinxia Li[1,2], Yaping Yan[1,2], Lan Dong[1,2], Jie Guo[1,2], Adili Aiyiti[1,2], Xiangfan Xu[1,2,3][†], Baowen Li[4][‡]

[1] Center for Phononics and Thermal Energy Science, School of Physics Science and Engineering, Tongji University, Shanghai 200092, China
[2] China-EU Joint Lab for Nanophononics, Tongji University, Shanghai 200092, China
[3] Institute for Advanced Study, Tongji University, Shanghai 200092, China
[4] Department of Mechanical Engineering, University of Colorado, Boulder, CO 80309, USA

Correspondence and requests for materials should be addressed to
X.X.(xuxiangfan@tongji.edu.cn) or B.L.(Baowen.Li@Colorado.edu)

**Abstract**
The needs for efficient heat removal and superior thermal conduction in nano/micro devices have triggered tremendous studies in low-dimensional materials with high thermal conductivity. Hexagonal boron nitride (*h*-BN) is believed to be one of the candidates for thermal management and heat dissipation due to its novel physical properties, i.e. thermal conductor and electrical insulator. Here we reported interfacial thermal resistance between few-layer *h*-BN and its silicon oxide substrate using differential 3ω method. The measured interfacial thermal resistance is around $\sim 1.6 \times 10^{-8}$ m$^2$K/W for monolayer *h*-BN and $\sim 3.4 \times 10^{-8}$ m$^2$K/W for 12.8nm-thick *h*-BN in metal/*h*-BN/SiO$_2$ interfaces. Our results suggest that the voids and gaps between substrate and thick *h*-BN flakes limit the interfacial thermal conduction. This work provides a deeper understanding of utilizing *h*-BN flake as lateral heat spreader in electronic and optoelectronic nano/micro devices with further miniaturization and integration.

## Introduction

Recent advances in nanotechnology and material science enable us to fabricate electronic and optoelectronic devices down to nano/micro scale. With further miniaturization of modern electronics, the accumulated high-dense waste heat can rise the operating temperature into a high level where heat dissipation and thermal management become crucial for the performance and stability of the devices [1]. To this end, nanomaterials with superior thermal conductivity ($\kappa$) are proposed and attract condensed studies in the last few decades. Two-dimensional (2D) materials, e.g. graphene (with $\kappa$ up to $\sim$5300 Wm$^{-1}$K$^{-1}$) and hexagonal boron nitride *h*-BN (with $\kappa$ up to $\sim$200 Wm$^{-1}$K$^{-1}$ to $\sim$500 Wm$^{-1}$K$^{-1}$), are believed to be potential candidates for efficient heat removal due to their high in-plane thermal conductivity [2-11].

Single layer graphene holds superior thermal conductivity due to the light element weight of carbon atoms and the strong bonding between them. Its thermal conductivity increases with thickness and approaches that in bulk graphite [11, 12], making supported multi-layer graphene a prospective material for thermal management applications in nanoelectronic circuits. Due to the geometric similarity, *h*-BN also holds a high thermal conductivity [2, 13, 14]. However and more importantly, despite the large in-plane thermal conductivity, the interfacial thermal resistance between 2D materials and substrate dominates the thermal conduction, which plays a critical role in performance and current saturation of field-effect transistors.

A number of heterostructures of stacked 2D materials, such as graphene/$h$-BN, MoS$_2$/ $h$-BN and black phosphorous/$h$-BN, have emerged recently [15]. The in-plane electrical transport of graphene in graphene/$h$-BN heterostructure has much higher mobility than that in graphene/SiO$_2$, and reaches an ultra high value of 1,000,000 cm$^2$V$^{-1}$s$^{-1}$ at low temperature [16, 17]. In this type of 2D structures, the applied large in-plane electrical voltage could result in local heating and hot spot, making the thermal management become a bottleneck for device performance.

Many attentions have been paid to study the role of phonon-phonon interaction in heat conduction between graphene and the related dielectric substrate, comparing to the rare study in the interface between $h$-BN and the dielectric substrate. In this paper, we studied the interfacial thermal resistance of metal/$h$-BN/SiO$_2$ heterostructure using 3ω method. The measured interfacial thermal resistance in metal/$h$-BN/SiO$_2$ interface increases from ~1.6×10$^{-8}$ m$^2$K/W to ~ 3.4×10$^{-8}$ m$^2$K/W when the thickness of $h$-BN layer changes from ~1nm (respects to monolayer $h$-BN) to 12.8nm. We also find that the voids and gaps between $h$-BN and SiO$_2$ interface increase in thicker $h$-BN layers and dominate the total thermal resistance in metal/$h$-BN/SiO$_2$ heterostructure.

## Materials and Method

Before measuring the interfacial thermal resistance in metal/$h$-BN/SiO$_2$ heterostructure, we utilize standard 3ω method to measure the thermal conductivity of silicon oxide thin film and the silicon substrate underneath, to test the validation of 3ω method and estimate its measurement uncertainty or error bars.

*Measuring thermal conductivity of silicon wafer using 3ω method*

A thin Cr/Au (5nm/50nm) electrode, with 3μm in width and 33μm in length, was firstly deposited onto SiO$_2$/Si wafer using E-beam lithography and thermal evaporation. An AC-current with frequency of 1ω was applied into the electrode, introducing a temperature fluctuation with frequency of 2ω. The resistance of the electrode can be treated to have a linear dependence with temperature when the temperature rise is small enough. Consequently, an AC voltage with frequency of 3ω can be detected in the electrode, whose amplitude vs. frequency is related to how much heat dissipating into the substrate, i.e. the thermal conduction of the substrate. When the heat penetration depth λ is much larger than the half width of the electrode but smaller than the thickness of silicon substrate, the thermal conductivity of silicon substrate can be simplified into [18, 19]:

$$\kappa_{bulk} = \frac{V^3 \ln(\omega_2/\omega_1)}{4l\pi R^2 \cdot (V_{3\omega,1} - V_{3\omega,2})} \frac{dR}{dT} ,  \qquad (1)$$

where $\kappa_{bulk}$ is the thermal conductivity of silicon, $R$ is the resistance of electrode, $l$ is the length of electrode, $V$ is the voltage with frequency ω, $V_{3\omega,1}$ and $V_{3\omega,2}$ are the measured 3ω-voltage at two different frequencies, i.e. $\omega_1$ and $\omega_2$, respectively. $T_{2\omega}$ is the temperature rise in electrode, which can be calculated from:

$$T_{2\omega} = 2 \frac{dT}{dR} \frac{R}{V} V_{3\omega}. \qquad (2)$$

To test the validation of the measurement setup of 3ω method, we firstly measured thermal conductivity of silicon wafers. Figure 1a shows the measured resistance of electrode versus temperature. In the measured temperature range, i.e. 50K to 300K, electrode resistance $R$ increases linearly with temperature, following $R=3.46+0.0165\,T$. The measured $T_{2\omega}$ as a function of ln$\omega$ is shown in Figure 1b, the slope of which is related to the thermal conductivity of silicon substrate. At $T=300$K, the thermal conductivity was measured to be around ~127 Wm$^{-1}$K$^{-1}$, slightly lower than that in clean silicon (~140 Wm$^{-1}$K$^{-1}$). This is reasonable since thermal conductivity in silicon decreases with doping level, due to the phonon-impurity scatterings. As shown in Figure 1c, thermal conductivity of low-doped silicon (with electrical conductivity of $5\times10^{-4}$ S/cm, regarded as clean silicon) is around ~140 Wm$^{-1}$K$^{-1}$ at $T=300$K, which reduces into ~106 Wm$^{-1}$K$^{-1}$, ~71.2 Wm$^{-1}$K$^{-1}$, ~45 Wm$^{-1}$K$^{-1}$, ~40.7 Wm$^{-1}$K$^{-1}$ in samples with electrical conductivity of ~$3.6\times10^2$ S/cm, ~$1.5\times10^3$ S/cm, ~$2.2\times10^3$ S/cm, ~$3.8\times10^3$ S/cm [20], respectively. Our measured thermal conductivity of the silicon substrate with electrical conductivity of ~$1\times10^2$ S/cm is consistent with the existing results reported by other groups (Figure 1c).

*Measuring thermal conductivity of SiO$_2$ thin film using differential 3ω method*

In this section, the design and setup of differential 3ω method for thin film thermal conductivity measurement are described. The same method was used to measure interfacial thermal resistance of metal/$h$-BN/SiO$_2$ interface, which will be shown later. Here, two SiO$_2$/Si samples with different thickness of SiO$_2$ are used. A thin Cr/Au (5nm/50nm) electrode, with 3μm in width and 33μm in length, was deposited onto SiO$_2$/Si wafers with the SiO$_2$ thickness chosen to be 100 nm and 300 nm, respectively. An AC-current was applied into electrode and introduces a temperature rise in the electrode, but with different amplitude, i.e. $T_{2\omega(100\text{nm})}$ and $T_{2\omega(300\text{nm})}$ for 100nm-thick SiO$_2$ and 300nm-thick SiO$_2$, respectively. When heat penetration depth λ is smaller than the thickness of silicon wafer, and half width of electrode is much larger than thickness of SiO$_2$, the thermal conductivity of SiO$_2$ can be calculated from [21-24]:

$$\kappa = \frac{P\cdot t}{\Delta T_{2w}\cdot S}, \qquad (3)$$

where $\Delta T_{2\omega}= T_{2\omega(300\text{nm})} - T_{2\omega(100\text{nm})}$, $t$ is the thickness of SiO$_2$ film and $S$ is the cross section area of SiO$_2$ under the electrode. Figure 1b shows the measured temperature $T_{2\omega}$ versus ln$\omega$ of 100nm-thick SiO$_2$/Si ($T_{2\omega(100\text{nm})}$) and 300nm-thick SiO$_2$/Si ($T_{2\omega(300\text{nm})}$). Since temperature scales linearly with ln$\omega$ when the frequency is in the range between 200 Hz and 2000 Hz, it is safe to calculate $\Delta T_{2\omega}$ using the data when ω=1000Hz. The measured room temperature thermal conductivity of SiO$_2$ is around 1.1±0.1 Wm$^{-1}$K$^{-1}$, which is slightly low than that in bulk SiO$_2$ due to the phonon-surface scatterings (Figure 1d) [25]. The thermal conductivity decreases with decreasing temperature with no visible peak at lower temperature, which is the fingerprint of amorphous materials.

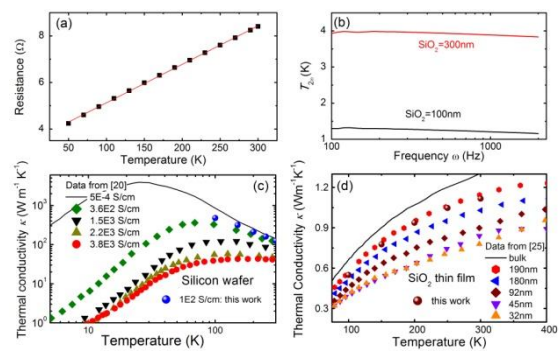

Figure 1 Measuring thermal conductivity of Silicon substrate and SiO$_2$ film by 3ω method. (a) Resistance of Cr/Au electrode versus temperature. (b) Temperature changes versus the frequency of AC-current source. (c) The measured thermal conductivity of silicon. The data (with different electrical conductivity) from reference [20] are listed for comparation. (d) The measured thermal conductivity of SiO$_2$ film. The data from reference [25] are listed for comparation.

## Fabrication of metal/$h$-BN/SiO$_2$ interface and the measurement details

The sample fabrication process is summarized in Figure 2. Boron nitride flake with size larger than 30μm was deposited onto 300nm-thick SiO$_2$/Si wafer by exfoliation using scotch tape method (Figure 2a & 2b). Next, the samples were annealed in mixed gas of 5％H$_2$/95％Ar with flow rate of 400ml/min at 400 ℃ for three hours to remove possible organic residues and to improve the interface, followed by standard E-beam lithography and lift-off process (Figure 2c & 2d). To facilitate the differential 3ω measurement, two identical electrodes (or heaters) were fabricated onto the sample wafer, namely, one on the surface of $h$-BN flake and the other one directly on SiO$_2$ (the reference electrode in Figure 2c). At the final step, O$_2$ plasma was used to pattern $h$-BN flakes to match the shape of the metal electrode (Figure 2e & 2f), with the purpose to eliminate heat dissipation in lateral direction and to ensure that heat flow in one dimension from electrode to the substrate.

The measurement was conducted in the temperature ranging from $T$=50K to $T$=300K in a Variable Temperature Instrument (Janis ST-400 VTI) with vacuum better than $1\times10^{-4}$ pa. The AC-current frequency was swept from 200Hz to 10000Hz (Figure 3a) and the temperature difference $\Delta T_{2\omega}$ was taken at ω=1000Hz. We have confirmed that the $\Delta T_{2\omega}$ keeps a constant value in the full sweeping range. For example, the contact resistance taken at 200Hz, 1000Hz and 2500Hz fluctuate within 1.5 % at $T$=300K.

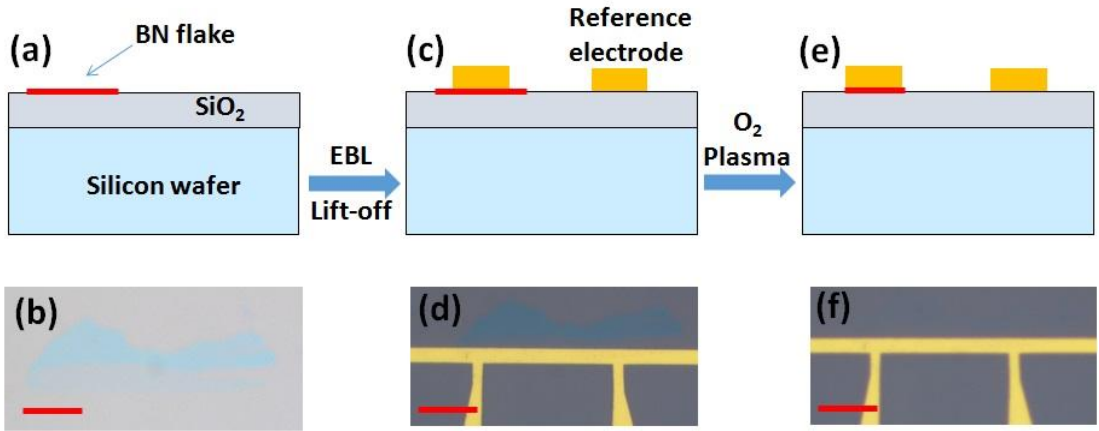

Figure 2 Fabrication of metal/$h$-BN/SiO$_2$ interface suitable for thermal resistance measurement. (a)&(b) Few-layer $h$-BN transferred to SiO$_2$/Si substrate by scotch-tape mediated exfoliation. The scale bar is 20μm. (c)&(d) Cr/Au electrode deposited on the SiO$_2$/Si substrate through e-beam lithography and thermal evaporation technique. The scale bar is 15μm. (e)&(f) O$_2$ plasma is used to pattern $h$-BN thin flake, the scale bar is 15μm.

In this sample geometry, the measured thermal resistance is consisted of the contribution from metal/$h$-BN interface, cross-plane resistance of $h$-BN flake, $h$-BN/SiO$_2$ interface and metal/SiO$_2$ interface (the reference electrode). Therefore, the measured resistance $R_{th}$ can be written as $R_{th}=R_{metal/BN} + R_{BN} + R_{BN/SiO2} - R_{metal/SiO2}$. The cross-plane thermal resistance of few-layer $h$-BN sample is in the order of $10^{-9}$ m$^2$KW$^{-1}$ (assuming c-axis $\kappa$~1.5 Wm$^{-1}$K$^{-1}$ in few-layer $h$-BN [26]), which is much lower than the measured thermal resistance for few layer $h$-BN, especially when the thickness of $h$-BN flake is thinner than 3.7nm.

## Results and Discussions

Four metal/$h$-BN/SiO$_2$ interfaces with different thickness (1nm, 1.5nm, 3.7nm and 12.8nm) were measured. Based on the Atomic Force Microscope measurement and optical microscope, the $h$-BN samples with 1nm and 1.5nm in thickness are believed to be monolayer and bilayer (or tri-layer), respectively. The interfacial thermal resistance of these four samples demonstrate weakly temperature dependence and increase slightly with decrease of temperature (Figure 3b), similar to previous observation on that in graphene/SiO$_2$ interface (stars in Figure 3b). The thermal resistance in monolayer metal/$h$-BN/SiO$_2$ interface is measured to be around ~ $1.6 \times 10^{-8}$ m$^2$K/W at $T$=300K and ~ $2.6 \times 10^{-8}$ m$^2$K/W at $T$=50K, which is relatively larger than that in monolayer graphene/SiO$_2$ interface measured by the same 3ω method. Nevertheless, the room temperature value is consistent with that calculated by MD simulation (~$2.2 \times 10^{-8}$ m$^2$K/W) of $h$-BN/SiO$_2$ interface [27].

The thermal resistance of $R_{metal/BN}$ and $R_{metal/SiO2}$ are independent with the thickness of $h$-BN flake, therefore the observed thickness-dependent thermal resistance $R_{th}$ at $T$=300K (Figure 4) should be contributed from both $R_{BN/SiO2}$ and $R_{BN}$. However, the cross-plane $R_{BN}$ is in the order of $10^{-9}$ m$^2$KW$^{-1}$, which is much smaller than that in $R_{th}$ and its grow rate with thickness, we can safely assume that $R_{BN/SiO2}$ have major contribution to the increase of measured $R_{th}$ (especially for $h$-BN thinner than 3.7nm). For thicker samples, i.e. thickness >3.7nm, on the other hand, both $R_{BN/SiO2}$ and $R_{BN}$ contribute to the increase of measured $R_{th}$. It is more possible that $R_{BN}$ dominates the increase of measured $R_{th}$ for samples with $h$-BN thicker than 3.7nm, since the increase of $R_{th}$ and $R_{BN}$ are parallel. It is worth to point out that, from sample with $h$-BN thickness of 1nm, the estimated the thermal resistance of $R_{metal/BN}$ should be in the order of ~$1 \times 10^{-8}$ m$^2$KW$^{-1}$ when considering the fact that $R_{metal/SiO2}$ are reported to be between ~$5 \times 10^{-9}$ m$^2$KW$^{-1}$ and ~$1 \times 10^{-8}$ m$^2$KW$^{-1}$ [28-30] and that cross-plane thermal resistance of few-layer $h$-BN has minor contribution to $R_{th}$.

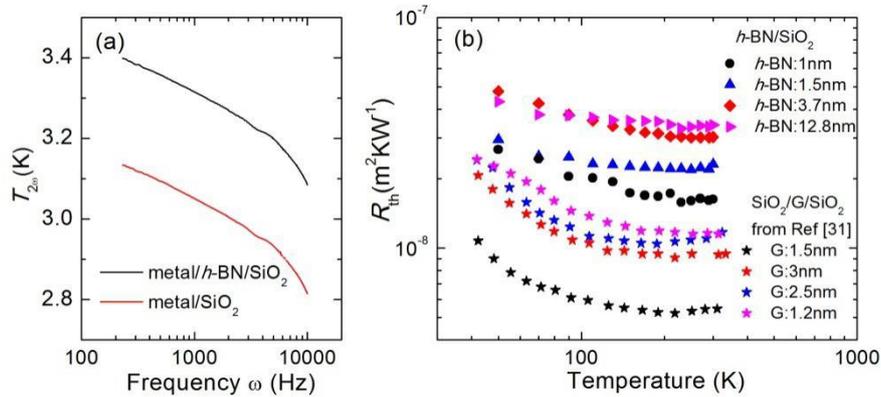

Figure 3. (a) Temperature changes versus the frequency of AC-current source for 1.5nm-thick $h$-BN samples. (b) The measured interfacial thermal resistance as a function of temperature. The reported data on SiO$_2$/Graphene/SiO$_2$ interface from reference [31] are listed for comparation.

The question whether interfacial thermal resistance of 2D materials and SiO$_2$ substrate changes with the thickness of 2D materials attracts condensed experiments or simulation studies. Ni *et al*. examined few-layer $h$-BN encased by SiO$_2$ using equilibrium MD simulation and found that the interfacial thermal resistance is independent with thickness of $h$-BN on amorphous SiO$_2$, but dependent with thickness of $h$-BN on crystal SiO$_2$ [27]. Chen *et al*. carried out SiO$_2$/Graphene/SiO$_2$ interfacial thermal resistance measurement using 3ω method and found no clear dependence on the sample thickness [31].

Koh *et al*. found that the interfacial thermal resistance on $SiO_2$/Graphene interface is around ~ $4 \times 10^{-8}$ $m^2K/W$ irrespective of graphene layers using TDTR method (Time-domain Thermo-reflectance) [28]. This is consistent with our results in thicker samples (from 3.7nm to 12.8nm) that the thermal resistance is thickness independent (the slight increase is probably due to the cross-plane thermal resistance in *h*-BN). More interestingly, we observed a strong thickness-dependent thermal resistance in the measured temperature between 50K and 300K with *h*-BN thickness changing from 1nm to 3.7nm. The measured room temperature interfacial thermal resistance increase from ~$1.6 \times 10^{-8}$ $m^2K/W$ for monolayer metal/*h*-BN/$SiO_2$ interface to ~ $3 \times 10^{-8}$ $m^2K/W$ for 3.7nm-thick metal/*h*-BN/$SiO_2$. This thickness dependence could not be explained by the increase of cross-plane thermal resistance through *h*-BN layers, as the cross-plane thermal resistance of 3.7nm *h*-BN flakes is around ~$2.5 \times 10^{-9}$ $m^2K/W$ when assuming the cross-plane thermal conductivity of *h*-BN to be around 1.5 $Wm^{-1}K^{-1}$ [26]. Furthermore, the cross-plane thermal resistance could increase linearly with the thickness of *h*-BN, which however is different from what observed in Figure 4.

Alternatively, the observed thickness dependence should be related to the roughness of interface between *h*-BN and $SiO_2$. Atomic Force Microscope (AFM) shown in Figure 5 provides evidence to support the assumption of the roughness changes. The root mean square (RMS) of 500nm×500nm area in 1nm-thick *h*-BN top surface is around 0.135nm (measurement was carried out before EBL and thermal evaporation), which is closed to that measured in $SiO_2$ substrate (~0.175nm). The measurement indicates that monolayer *h*-BN flake conforms to the surface of $SiO_2$ substrate either during exfoliation or annealing process. The measured RMS of 1.5 nm and 3.7nm *h*-BN flakes decreases to 0.092nm and 0.080nm, respectively. The roughness on top surface of *h*-BN decreases in thicker *h*-BN flakes due to the enhancement of the stiffness, leaving more voids and gaps between *h*-BN and the $SiO_2$ underneath. Although the AFM only measure the topography of *h*-BN top surface, it reflects indirectly the roughness of the interface underneath, and indicates that the gaps, voids and wrinkles dominate the interfacial thermal resistance in metal/*h*-BN/$SiO_2$ interface. The picture of roughness induced thermal resistance is also supported by the fact that the measured interfacial thermal resistance is much larger than that in *h*-BN/$SiN_x$ interface which was fabricated by PMMA-mediated wet transfer method [13].

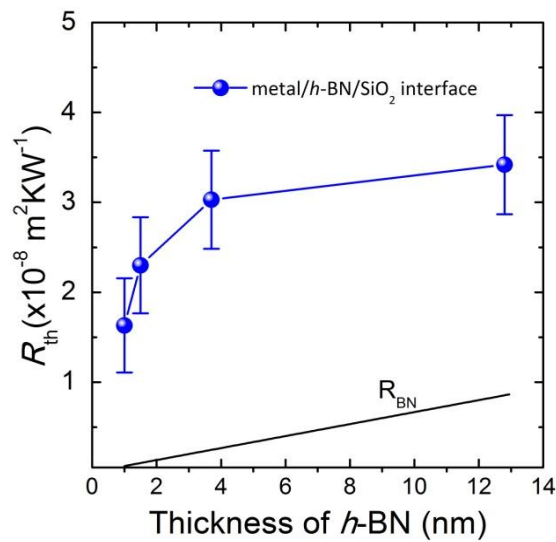

Figure 4. Room temperature interfacial thermal resistance with respect to the thickness of *h*-BN layers

Boron nitride flakes are proposed to be a candidate for potential applications in thermal dissipation and heat removal due to its superior in-plane thermal conductivity which is believed to be higher than that in most of metals [2]. However, the interfacial thermal resistance between 2D materials and substrate plays a dominate role in heat dissipation due to further miniaturization of modern electronics where billions of transistors integrate in every square centimetres. Our experimental result suggests that $h$-BN flakes are rather a thermal insulator than a thermal conductor due to its high interfacial thermal resistance in nanoscale. From the measured data, we can estimate the Kapitza length ($l_k$) as $l_k=\kappa R_k$, where $R_k$ is the Kapitza resistance. Using $\kappa=1.1 \text{Wm}^{-1}\text{K}^{-1}$ and $R_k=\sim 1.6\times 10^{-8}$ m$^2$K/W for the measured SiO$_2$ and monolayer $h$-BN/SiO$_2$ interface, $l_k$ can be estimated to be around 17.6nm. This indicates that the sub-nanoscale interface of monolayer $h$-BN and SiO$_2$ provides the same resistance of 17.6nm-thick SiO$_2$, considering the fact that SiO$_2$ is a good thermal insulator.

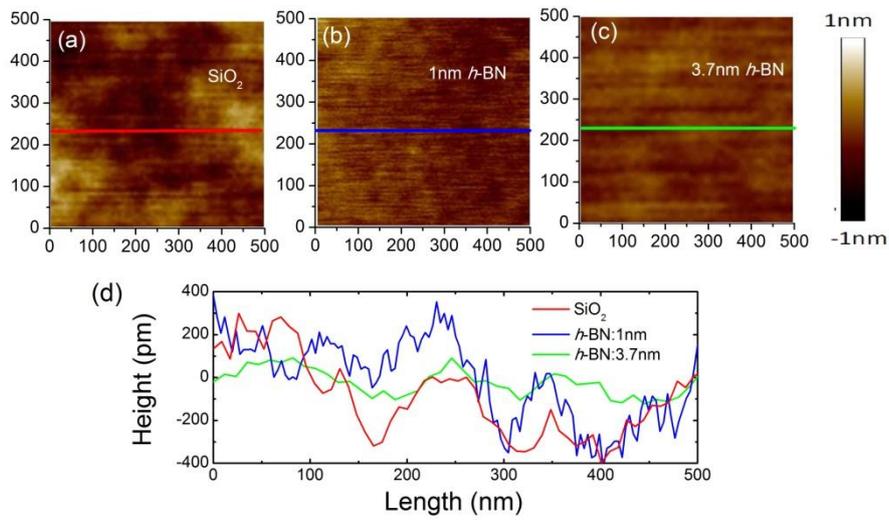

Figure 5. The Atomic Force Microscope results on the roughness of the SiO$_2$ surface, 1nm-thick $h$-BN surface and 3.7nm-thick $h$-BN surface, respectively.

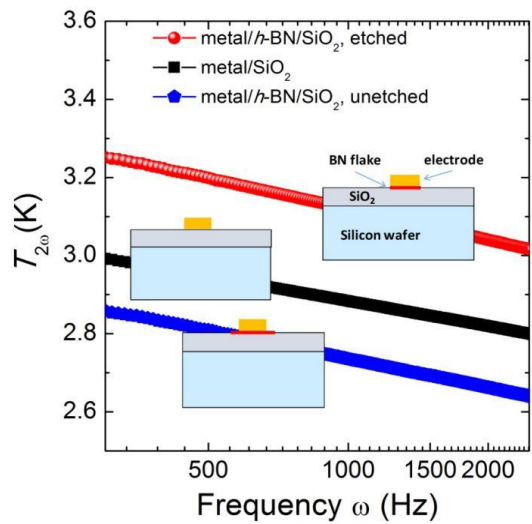

Figure 6. Temperature rise in metal/$h$-BN/SiO$_2$ interface with and without O$_2$ plasma pattern.

Therefore, despite the high thermal conductivity in h-BN flake, the interface in h-BN/SiO$_2$ or 2D heterostructure [32] is a thermal insulator. Figure 6 demonstrates the $T_{2\omega}$ versus frequency in metal/h-BN/SiO$_2$ interface with and without O$_2$ plasma pattern. The measured $T_{2\omega}$ in un-etched metal/h-BN/SiO$_2$ interface is much lower than that in etched metal/h-BN/SiO$_2$ interface and pure metal/SiO$_2$ interface, indicating that in-plane thermal conduction dominates the heat flow in h-BN/SiO$_2$ interface when the size of h-BN is larger than that in metal electrode on the top.

To further understand the heat flow across h-BN and SiO$_2$ interface and its effect on heat dissipation in 2D material based nanoelectronic devices, we brainstorm two imaginary 5nm-thickness h-BN based heterostructures: MicroDevice A, with length $l$=2μm and width $w$=500nm, and NanoDevice B, with length $l$=50nm and width $w$=20nm. The estimated in-plane and cross-plane thermal resistance is around ~1.3×10$^6$ K/W and ~3×10$^4$ K/W for MicroDevice A, indicating the cross-plane heat flow dominates the thermal dissipation. However for NanoDevice B, the estimated in-plane and cross-plane thermal resistance is around ~8×10$^5$ K/W and ~3.3×10$^7$ K/W, indicating in-plane heat conduction plays the critical role. This result emphasizes the importance of the metal contact in dissipating the waste thermal energy and in maintaining the nanoelectronic device for stable performance.

**Conclusion**

We carried out interfacial thermal resistance measurement in metal/h-BN/SiO$_2$ interface using differential 3ω method. Four devices with different thickness of h-BN layer, i.e. 1nm, 1.5nm, 3.7nm and 12.8nm, were tested. The measured interfacial thermal resistance increases with increased h-BN thickness, which is inconsistent with previous studies in graphene/SiO$_2$. We attribute this thickness dependence to the voids and gaps in h-BN/SiO$_2$ interface, which was introduced during exfoliation process. This picture is supported by the AFM results that surface roughness of h-BN flakes reduces in thicker h-BN samples. The room temperature thermal resistance is around ~ 1.6×10$^{-8}$ m$^2$K/W for monolayer h-BN and ~ 3×10$^{-8}$ m$^2$K/W for 3.7nm-thick h-BN in metal/h-BN/SiO$_2$ interface, suggesting the dominate heat flow through h-BN/SiO$_2$ interface in micron scale 2D heterostructure devices.

**Acknowledgements**

This work was supported by National Natural Science Foundation of China (No. 11674245 & No. 11304227 & No. 11334007) and by the Fundamental Research Funds for the Central Universities (No. 2013KJ024).